\documentclass[letterpaper]{article} 
\usepackage{aaai25}  
\usepackage{times}  
\usepackage{helvet}  
\usepackage{courier}  
\usepackage[hyphens]{url}  
\usepackage{graphicx} 
\usepackage{natbib}  
\usepackage{caption} 
\usepackage{algorithm}
\usepackage{graphicx}
\usepackage{multirow}
\usepackage{booktabs}
\usepackage{enumitem}
\usepackage{color}
\usepackage[shortcuts]{extdash}
\usepackage[shortcuts]{extdash}
\usepackage{float}
\usepackage{tabularx}
\usepackage{mathbbol}
\usepackage{amsmath}
\usepackage{amssymb}
\usepackage[table]{xcolor}
\usepackage{colortbl}

\usepackage{newfloat}
\usepackage{listings}
\DeclareCaptionStyle{ruled}{labelfont=normalfont,labelsep=colon,strut=off} 
\floatstyle{ruled}
\newfloat{listing}{tb}{lst}{}
\floatname{listing}{Listing}

\title{Boosting ViT-based MRI Reconstruction from the Perspectives of \\ Frequency Modulation, Spatial Purification, and Scale Diversification}
\author{
    Yucong Meng\textsuperscript{\rm 1 2}\equalcontrib, Zhiwei Yang\textsuperscript{\rm 1 2 3}\equalcontrib, Yonghong Shi\textsuperscript{\rm 1 2}\thanks{Corresponding author.}, Zhijian Song\textsuperscript{\rm 1 2}\footnotemark[2]
}
\affiliations{
    \textsuperscript{\rm 1}Digital Medical Research Center, School of Basic Medical Science, Fudan University, Shanghai 200032, China\\
    \textsuperscript{\rm 2}Shanghai Key Laboratory of Medical Image Computing and Computer Assisted Intervention, Shanghai 200032, China\\
    \textsuperscript{\rm 3}Academy for Engineering and Technology, Fudan University, Shanghai 200433, China\\
    \{ ycmeng21, zwyang21 \} @m.fudan.edu.cn, \{ yonghong.shi, zjsong \} @fudan.edu.cn
}

\begin{document}

\maketitle
\begin{abstract}
The accelerated MRI reconstruction process presents a challenging ill-posed inverse problem due to the extensive under-sampling in k-space. Recently, Vision Transformers (ViTs) have become the mainstream for this task, demonstrating substantial performance improvements. However, there are still three significant issues remain unaddressed: (1) ViTs struggle to capture high-frequency components of images, limiting their ability to detect local textures and edge information, thereby impeding MRI restoration; (2) Previous methods calculate multi-head self-attention (MSA) among both related and unrelated tokens in content, introducing noise and significantly increasing computational burden; (3) The naive feed-forward network in ViTs cannot model the multi-scale information that is important for image restoration. In this paper, we propose FPS-Former, a powerful ViT-based framework, to address these issues from the perspectives of frequency modulation, spatial purification, and scale diversification. Specifically, for issue (1), we introduce a frequency modulation attention module to enhance the self-attention map by adaptively re-calibrating the frequency information in a Laplacian pyramid. For issue (2), we customize a spatial purification attention module to capture interactions among closely related tokens, thereby reducing redundant or irrelevant feature representations. For issue (3), we propose an efficient feed-forward network based on a hybrid-scale fusion strategy. Comprehensive experiments conducted on three public datasets show that our FPS-Former outperforms state-of-the-art methods while requiring lower computational costs.
\end{abstract}

\section{Introduction}
Magnetic Resonance Imaging (MRI) offers advantages such as non-radiation, high resolution, and superior contrast, making it essential in clinical diagnostics \cite{yang2022model,chen2022ai}. However, its long scanning times often increase the physical burden on patients. Additionally, involuntary movements like breathing, swallowing, and heartbeats usually blur images, limiting MRI's application. Reducing k-space acquisition can speed up MRI with fewer constraints than hardware modifications, but such undersampling introduces artifacts per the Nyquist theorem \cite{zeng2021review}. Eliminating these artifacts and reconstructing high-quality MRI images remains a significant challenge.

In recent years, many methods have adopted various CNN architectures for MRI reconstruction \cite{zeng2020comparative,aghabiglou2021mr,aghabiglou2021projection}. Due to the powerful non-linearity and feature representation capabilities of CNNs, CNN-MRI outperforms traditional compressed sensing (CS) based methods \cite{tamir2016generalized}. However, the convolutional operation has intrinsic characteristics such as local receptive fields and independence of input content \cite{li2021survey,zheng2022attention}. Therefore, CNN-based models cannot eliminate long-range degradation perturbation and gain suboptimal MRI reconstruction performance \cite{zhou2020dudornet,knoll2020deep}. 

\begin{figure}[t]
    \centering
    \includegraphics[width=1.0\linewidth]{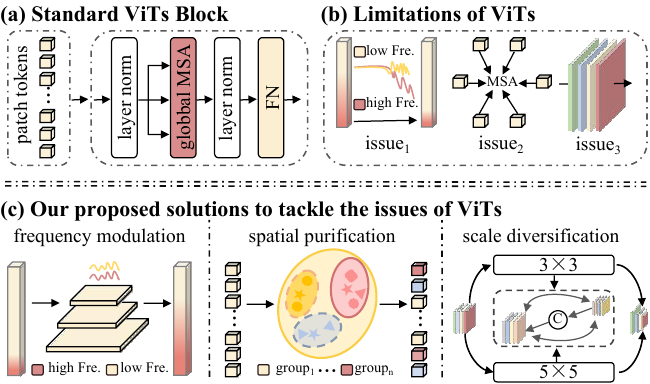}
    \caption{Our main idea. (a) The pipeline of standard ViTs block. (b) ViTs suffer limitations of high-frequency attenuation, irrelevant token interactions, and a lack of multi-scale feature representation. (3) We propose to tackle the above issues from the perspectives of frequency modulation, spatial purification, and scale diversification, thereby enhancing the performance of ViT-based MRI reconstruction.
}
    \label{fig:Figure1}
\vspace{-1.5 em}
\end{figure}
To alleviate such limitations, Vision Transformers (ViTs) \cite{dosovitskiy2020image} have been applied, shedding new light on MRI reconstruction tasks \cite{huang2022swin,10251064}. As shown in Figure~\ref{fig:Figure1} (a), ViTs stack Multi-head Self-Attention (MSA) blocks, treating each image patch as a semantic token and modeling their interactions globally \cite{9716741}. Unlike CNNs, which hierarchically enlarge the receptive field from local to global, even a shallow ViT can effectively capture global contexts, resulting in highly competitive performance for various computer vision tasks \cite{ali2023vision,Yang2024TacklingAF,Zhao_2023_CVPR}. 

However, ViTs still struggle to restore MRI details, facing several critical issues as shown in Figure~\ref{fig:Figure1} (b): (1) ViTs are limited in capturing high-frequency information, impairing their ability to detect local textures and edges essential for effective MRI reconstruction. As demonstrated in \cite{park2022vision,wang2022anti}, the MSA inherently amounts to a low-pass filter, which indicates that ViTs will overlook high-frequency information crucial for image restoration when it scales up its depth. (2) Standard ViTs calculate MSA among both related and unrelated tokens, introducing noise and increasing computational burden. Previous ViT-based methods linearly project all patch tokens into query, key, and value, and then perform matrix multiplication for MSA \cite{Zhou_2023_WACV,shen2024magneticresonanceimageprocessing}. However, some patches in MRI images are not related in content. Handling all tokens simultaneously introduces content-irrelevant noise and significantly increases computational complexity. (3) The multi-scale representation provides complementary information and plays a vital role in MRI reconstruction, while MSA in standard ViTs fails to effectively model multi-scale features \cite{Chen_2021_ICCV,Cai_2023_ICCV}.

By addressing the aforementioned issues from the perspectives of Frequency modulation, spatial Purification, and Scale diversification, we propose FPS-Former, a powerful ViT-based framework that significantly enhances the performance of MRI reconstruction, as shown in Figure~\ref{fig:Figure1} (c). Specifically, for issue (1), we propose the Frequency Modulation Attention Module (FMAM). FMAM recalibrates features in a Laplacian pyramid, enabling the retrieval of high-frequency information. This approach suppresses the low-pass filtering characteristic of ViTs, allowing the retention of more high-frequency details, which is beneficial for restoring local textures and edges. For issue (2), we design the Spatial Purification Attention Module (SPAM). Instead of processing all projected tokens simultaneously as standard ViTs, SPAM clusters tokens into different groups by identifying similar elements that yield the maximum inner product. Tokens within each group are considered closely related in content. The MSA operation is then applied within each group, reducing the noise impact of content-irrelevant tokens and significantly lowering computational complexity. For issue (3), we introduce a Scale Diversification Feed-forward Network (SDFN) that explores multi-scale feature representation by inserting two multi-scale deep convolution paths during feature transmission. Finally, observing that undersampled MRI images exhibit various types and degrees of degradation artifacts, we incorporate Hybrid Experts Feature Refinement (HEFR) into our model. HEFR comprises several convolutional layers and provides collaborative refinement for MRI reconstruction.

The main contributions of our work are listed as follows:
\begin{itemize}
    \item We propose the Frequency Modulation Attention Module to enhance the self-attention map by recalibrating frequency information in a Laplacian pyramid, selectively strengthening the contributions of shape and texture features, thereby overcoming the low-pass filtering of ViTs. 
    \item We introduce the Spatial Purification Attention Module to capture interactions among closely related tokens, thereby reducing redundant or irrelevant feature representations for precise self-similarity capturing.
    \item We propose the Scale Diversification Feed-forward Network to effectively model multi-scale information.
    \item Extensive experiments on both single-coil and multi-coil datasets under various undersampling patterns show that our method outperforms state-of-the-art (SoTA) competitors while requiring lower computational costs.
\end{itemize}

\section{Related Work}

\subsection{CNN-based MRI Reconstruction}
MRI reconstruction techniques can enhance image quality with less dependency on physiology and hardware, making them more accessible for accelerated MRI. Recent advances in deep learning have spurred the development of CNN-based MRI reconstruction. CMRNet pioneered the application of deep learning in MRI reconstruction by creating an offline CNN to map zero-filled to fully-sampled MRI images \cite{wang2016accelerating}. D5C5 proposed a CNN cascade for dynamic cardiac MRI \cite{schlemper2017deep}. DuDoRNet incorporated T1 priors for simultaneous k-space and image restoration \cite{zhou2020dudornet}. Dual-OctConv learned multi-scale spatial-frequency features from both real and imaginary components for parallel MRI \cite{feng2021dual}. Despite these successes, CNNs exhibit a limited receptive field and struggle to model long-range dependencies. Therefore, CNNs are suboptimal for restoring various image regions and cannot achieve satisfactory reconstruction performance \cite{khan2020survey,sarvamangala2022convolutional}. 

\subsection{ViT-based MRI Reconstruction}
Vision Transformers (ViTs) treat images as sequences of patches and use self-attention to capture global context \cite{yang2024separate}. Compared to CNNs, ViTs have advantages such as capturing global patterns and have been used for MRI reconstruction. As demonstrated in \cite{lin2022vision}, a ViT tailored for image reconstruction can achieve performance comparable to U-net while providing higher throughput and reduced memory consumption. SLATER addressed unsupervised MRI reconstruction by using a cross-attention module to capture correlations between latent variables and image features \cite{korkmaz2022unsupervised}. SwinMR designed a parallel imaging coupled swin transformer-based model for fast CS-MRI \cite{huang2022swin}. ReconFormer incorporated a local pyramid and global columnar ViT structure to learn multi-scale features at any stage, enabling enhanced reconstruction performance \cite{10251064}.

However, these methods still failed to achieve precise MRI reconstruction because they overlooked inherent issues of ViTs such as loss of high-frequency information, interference among unrelated patches, and the inability to model multi-scale features. By addressing these issues from three perspectives respectively, we propose FPS-Former to boost the performance of ViT-based MRI reconstruction.

\begin{figure*}[t]
    \centering
    \includegraphics[width=1.0\linewidth]{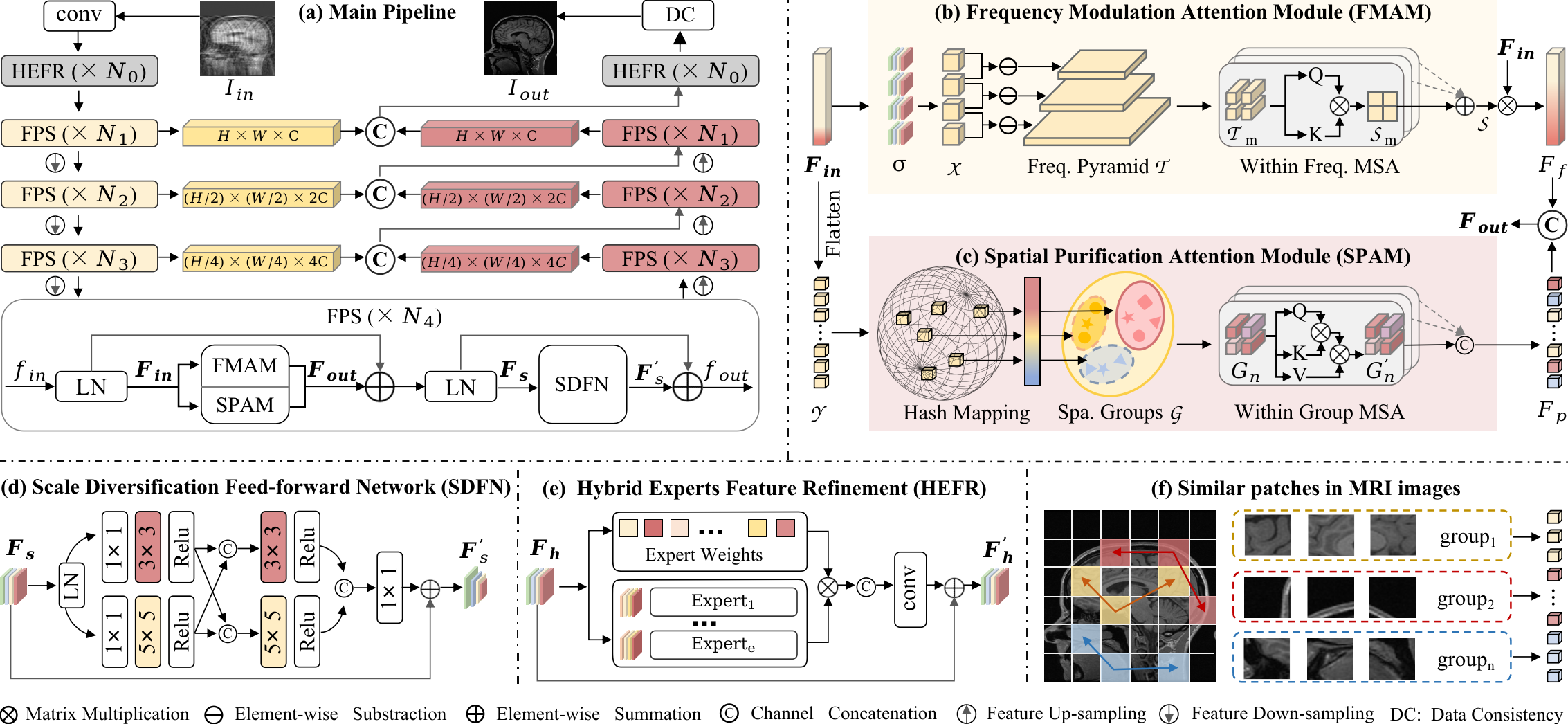}
    \caption{ (a) The overall architecture of the proposed FPS-Former. Given an input image $I_{in}$, we first apply a $3 \times 3$ convolution to obtain patch tokens. In the network backbone, we stack multiple FPS blocks to extract hierarchical features. FPS, consisting of FMAM (b), SPAM (c), and SDFN (d), is designed to tackle the issues of ViT-based MRI reconstruction. Besides, at both early and final stages of the network, we design HEFR (e) to provide refined features, ensuring the reconstruction of high-quality output $I_{out}$. (f) The motivation of our SPAM. MRI images contain widely distributed, similar patches that appear in groups.}
    \label{fig:Figure2}
    \vspace{-1.em}
\end{figure*}

\section{Methodology}

\subsection{Overall Pipeline}
As shown in Figure~\ref{fig:Figure2} (a), our proposed FPS-Former is a hierarchical encoder-decoder framework. Given a low-quality MRI image  $I_{in} \in \mathbb{R}^{H \times W \times C}$ with spatial resolution  $H \times W$ and channel dimension $C$, we first perform overlapped image patch embedding with a $3 \times 3$ convolution layer. Next, the embedding results are sent to the designed backbone, which stacks $N_{0}$ HEFR blocks and $N_{i\in \left [ 1,2,3,4 \right ] }$  FPS blocks. HEFR block is introduced to provide fine-grained information with multiple CNN-based experts, as shown in Figure~\ref{fig:Figure2} (e). The FPS block consists of the Frequency Modulation Attention Module (FMAM), Spatial Purification Attention Module (SPAM), and Scale Diversification Feed-forward Network (SDFN), as shown in Figure~\ref{fig:Figure2} (b), (c), and (d), respectively. It is designed to amend the mentioned issues of ViTs and extract hierarchical features with different spatial resolutions and channel dimensions. Then, the extracted features are sent to our decoder, which also includes $N_{i\in \left [ 1,2,3,4 \right ] }$ FPS and $N_{0}$ HEFR blocks. Skip connections are adopted to hierarchically bridge intermediate features between the encoder and decoder. Finally, a Data Consistency (DC) layer is added to reconstruct high-quality MRI image  $I_{out}$.

The above reconstruction process can be formulated as: $I_{out}=\mathcal{N}(I_{in})$, where $\mathcal{N}(\cdot )$ is the overall network and is trained by minimizing the following loss function:
\begin{equation}
\label{eq_1}
    \mathcal{L} =\left \| I_{out}- I_{gt}  \right \| _{1},
\end{equation}
where $I_{gt}$ denotes the ground-truth image, and $\left \| \cdot  \right \| _{1} $ is the $L_{1}$-norm. The proposed FPS and HEFR blocks will be specifically introduced in the following sections.

\subsection{FPS Block}
 ViTs in MRI reconstruction struggle with high-frequency information loss, irrelevant token interactions, and limited multi-scale feature modeling. To tackle these issues, we propose FPS block consisting of three main modules: Frequency Modulation Attention Module (FMAM), Spatial Purification Attention Module (SPAM), and Scale Diversification Feed-forward Network (SDFN). Formally, given the input feature $f_{in}$, the encoding produces of FPS is defined as:
\begin{equation}
f^{'}=  f_{in}+  {{\mathcal{F}}\divideontimes \mathcal{P}}({LN} \it(f_{in} )), f_{out}= f^{'}  +  \mathcal{S}(LN(\it f^{'}))
\end{equation}
where $LN(\cdot )$ denotes the layer normalization, $\mathcal{F}\divideontimes \mathcal{P}$ represents the combined effect of FMAM and SPAM, and $\mathcal{S}(\cdot )$ denotes the operation of SDFN.

\subsubsection{Frequency Modulation Attention Module}
Standard ViTs exhibit low-pass filtering characteristics, leading to the loss of high-frequency details such as texture for MRI reconstruction. To address this, we propose the Frequency Modulation Attention Module (FMAM) to recalibrate the importance of frequency at each level. Specifically, as shown in Figure~\ref{fig:Figure2} (b), we first use Gaussian functions with different variances to extract multiple Gaussian representations $\mathcal{X}$. The process can be formulated as follows:
\begin{equation}
\mathcal{X} = \left \{ \mathcal{X}_{m}  \right \} _{m=1}^{M+1}, \mathcal{X}_{m} =F_{in}\circledast \frac{1}{\sigma _{m} \sqrt{2\pi } } e^{- \frac{i^{2}+  j^{2}  }{2\sigma _{m}^{2} } }
\end{equation}
where the input feature $F_{in}\in \mathbb{R}^{H\times W\times C}$ is normalized from $f_{in}$. $(i,j)$ corresponds to the spatial location, $\sigma _{m\in\left [ 1,2,\dots ,M+1 \right ] }$ denotes the variance of the Gaussian function for the $m\-/th$ scale, and the symbol $\circledast$ represents the convolution operator. Then we construct the frequency pyramid $\mathcal{T}$
by subtracting adjacent elements in $\mathcal{X}$. This process is expressed as:
\begin{equation}
\mathcal{T} = \left \{ \mathcal{T}_{m}  \right \} _{m=1}^{M}, \mathcal{T}_{m} = \mathcal{X}_{m+1}-\mathcal{X}_{m} 
\end{equation}
The frequency pyramid $\mathcal{T}$ is composed of multiple layers, each containing distinct types of frequency information. To achieve a balanced distribution of low and high-frequency components within the model, we conduct Within Frequency MSA operation and effectively aggregate features from each frequency level. Specifically, we first calculate the attention scores $\mathcal{S}$ for each level of $\mathcal{T}$ as follows:
\begin{equation}
\mathcal{S} = \left \{ \mathcal{S}_{m}  \right \} _{m=1}^{M}, \mathcal{S}_{m} =  {\textstyle \sum_{i=1}^{I}} \operatorname{softmax}(({Q_{m}^{i} K_{m}^{i}})/{\sqrt{d}}) 
\end{equation}
where $I$ is the number of attention heads, $Q_{m}$ and $K_{m}$ are derived from $\mathcal{T}_{m}$ using linear transformations, and $d=(C/I$) denotes the dimension of each head. Finally, we sum the attention scores in $\mathcal{S}$ and multiply the result by the Value ($V$, derived from $F_{in}$) to obtain the result of FMAM $F_{f}$:
\begin{equation}
F_{f} = (  {\textstyle \sum_{m=1}^{M}(\mathcal{S}_{m} \in{\mathcal{S}})} )V 
\end{equation}

\subsubsection{Spatial Purification Attention Module}
As shown in Figure~\ref{fig:Figure2} (f), MRI images contain clusters of image patches that are similar within each group but distinctly different from those outside the group. Previous ViT-based methods perform a dense MSA operation on all patch tokens simultaneously. This operation leads to noisy interactions among unrelated features, hampering MRI reconstruction. To address this, we propose the Spatial Purification Attention Module (SPAM), which applies a sparsity constraint by computing self-attention only between contextually related tokens to reduce noise and computational complexity.

Specifically, as shown in Figure~\ref{fig:Figure2} (c), given the input feature map $F_{in}$, we first flatten it into $\smash{ \mathcal{Y} = \left \{ f_{j}\in \mathbb{R}^{C}  \right \} _{j=1}^{J}}$, where $J$ represents the number of tokens. Subsequently, we use the hash function to aggregate the information and map the C-dimensional tokens $f_{j}$ into integer hash codes $\mathcal{Z}$. This hash mapping can be formulated as:
\begin{equation}
\mathcal{Z}= \left \{\mathcal{Z}_{j} \in \mathbb{Z}  \right \} _{j=1}^{J}, \mathcal{Z}_{j} =\left \lfloor  (a \cdot f_{j} + b)/r \right \rfloor 
\end{equation}
where $a\in \mathbb{R}^{C}$ and $b\in \mathbb{R}$ are random variables satisfying $a\sim \mathcal{N}(0,1)$ and $b\sim \mathcal{U}(0,r)$, $r\in \mathbb{R}$ is a constant, $\lfloor \cdot \rfloor$ is the floor function. Next, we sort all elements in $\mathcal{Y}$ based on their hash code in $\mathcal{Z}$. The $j\-/th$ sorted element is denoted as $f_{j}^{'}$. Then we split them into groups $\mathcal{G}$, which is expressed as:
\begin{equation}
\mathcal{G} = \left \{ \mathcal{G}_{n}  \right \} _{n=1}^{N}, \mathcal{G}_{n} =\left \{ f_{j}^{'}: ng + 1\le j\le (n+1)g  \right \}
\end{equation}
where $N$ denotes the number of groups, and each group has $g$ elements. With such a scheme, closely related tokens are grouped together. Subsequently, we apply the Within Group MSA operation for each ${G}_{n}$ to obtain updated groups $\mathcal{G}^{'}$:
\begin{equation}
\begin{split}
\mathcal{G}^{'} = \left \{ G_{n}^{'}  \right \} _{n=1}^{N}, G^{'}_{n} = {\smash \sum_{i=1}^{I}}W_{i}head_{i}\left ({G}_{n} \right )
\end{split}
\end{equation}
where $head_{i}(\cdot)$ represents the self-attention operation of the $i\-/th$ head, $I$ is the number of attention heads, and $W_{i} \in R^{C \times d}$ represents the learnable parameters.

Next, we take out all the elements from each $G^{'}_{n}$ and unsort them according to their original positions in $\mathcal{Y}$. We then concatenate the elements to obtain the purified features $F_{p}$.

Finally, we concatenate $\left [ \cdot  \right ] $ the purified features $F_{p}$ with the frequency result $F_{f}$ from FMAM. A depthwise convolution $f(\cdot )$ is further applied to aggregate the information. In this way, we retain high-frequency information and achieve spatial purification. The above process is formulated as:
\begin{equation}
F_{out} = f(\left [ F_{p},F_{f}  \right ])  
\end{equation}

\subsubsection{Scale Diversification Feed-forward Network}
Multi-scale representations have been proven effective in enhancing MRI reconstruction. However, previous methods often focus on integrating single-scale components into feed-forward networks, overlooking the importance of multi-scale feature representations. To address this, we design a Scale Diversification Feed-forward Network (SDFN) by inserting two multi-scale depth-wise convolution paths in the transmission process as shown in Figure~\ref{fig:Figure2} (d). Specifically, given an input $F_{s}$, which is normalized from the above aggregated $F_{out}$, we first expand its channel dimension with $1 \times 1$ convolution in the ratio of $r$. Then, the obtained feature is sent into two parallel branches. During feature transformation, we use $3 \times 3$ and $5 \times 5$ depthwise convolutions to enhance multi-scale local information extraction. The entire feature fusion process of SDFN can be described as follows:
\begin{equation}
\vspace{-0.2 em}
\begin{aligned}
\begin{split}
&\hat{F_{s} } =f_{1\times 1}(LN(F_{s}) , \\
&F_{p_{1}} =\sigma (f_{3\times 3}(\hat{F_{s} }) ),F_{s_{1}} =\sigma (f_{5\times 5}(\hat{F_{s} })),\\
&F_{p_{2}} =\sigma (f_{3\times 3}\left [F_{p_{1}},F_{s_{1} } \right ]),
F_{s_{2} } =\sigma (f_{5\times 5}\left [F_{s_{1} },F_{p_{1} } \right ]),\\
&F_{s}^{'}=f_{1\times 1}\left [F_{p_{2} }, F_{s_{2} }\right ] + F_{s}
\end{split}
\end{aligned}
\end{equation}
where $f_{1\times 1}(\cdot )$ denotes the $1\times 1$ convolution, $\sigma(\cdot )$ is a ReLU activation, $f_{3\times 3}(\cdot )$ and $f_{5\times 5}(\cdot )$ denote $3\times 3$ and $5\times 5$ depth-wise convolutions, and $\left [ \cdot  \right ]$ is the  channel-wise concatenation.

\begin{table*}[t]
\renewcommand{\arraystretch}{0.8}
    \centering
    \small
    \setlength{\tabcolsep}{1.0mm}
\begin{tabular}{l|c|cccccc|cccccc}
\toprule
\multicolumn{1}{c|}{}                                  &                                               & \multicolumn{6}{c|}{CC359}                                                                                                                                                                  & \multicolumn{6}{c}{fastMRI}                                                                                                                                                                \\ \cmidrule{3-14} 
\multicolumn{1}{c|}{}                                  &                                               & \multicolumn{2}{c|}{NMSE $\downarrow$}                                             & \multicolumn{2}{c|}{SSIM $\uparrow$}                                             & \multicolumn{2}{c|}{PSNR $\uparrow$} & \multicolumn{2}{c|}{NMSE$\downarrow$}                                             & \multicolumn{2}{c|}{SSIM $\uparrow$}                                             & \multicolumn{2}{c}{PSNR $\uparrow$} \\ \cmidrule{3-14} 
\multicolumn{1}{l|}{\multirow{-3}{*}{Method}} & \multirow{-3}{*}{Type}               & AF=4   & \multicolumn{1}{c|}{AF=8}                           & AF=4   & \multicolumn{1}{c|}{AF=8}                           &AF=4    & AF=8   & AF=4   & \multicolumn{1}{c|}{AF=8}                           & AF=4   & \multicolumn{1}{c|}{AF=8}                           & AF=4   & AF=8   \\ \midrule
CS~\cite{tamir2016generalized}                                                     &                                            & 0.0483          & \multicolumn{1}{c|}{0.1066}                                  & 0.7510          & \multicolumn{1}{c|}{0.6424}                                  & 26.34            & 22.85           & 0.0583          & \multicolumn{1}{c|}{0.0903}                                  & 0.5736          & \multicolumn{1}{c|}{0.4870}                                  & 29.54           & 26.99           \\ \midrule
KIKI-Net~\cite{eo2018kiki}                                               &                                               & 0.0221          & \multicolumn{1}{c|}{0.0417}                                  & 0.8415          & \multicolumn{1}{c|}{0.7773}                                  & 28.97            & 26.24           & 0.0353          & \multicolumn{1}{c|}{0.0546}                                  & 0.7172          & \multicolumn{1}{c|}{0.6355}                                  & 31.87           & 29.27           \\
UNet-32~\cite{zbontar2018fastmri}                                                &                                               & 0.0197          & \multicolumn{1}{c|}{0.0385}                                  & 0.8898          & \multicolumn{1}{c|}{0.8348}                                  & 31.54            & 28.66           & 0.0337          & \multicolumn{1}{c|}{0.0477}                                  & 0.7248          & \multicolumn{1}{c|}{0.6570}                                  & 31.99           & 30.02           \\
D5C5~\cite{schlemper2017deep}                                                   &                                               & 0.0177          & \multicolumn{1}{c|}{0.0428}                                  & 0.8977          & \multicolumn{1}{c|}{0.8267}                                  & 31.59            & 28.20           & 0.0332          & \multicolumn{1}{c|}{0.0512}                                  & 0.7256          & \multicolumn{1}{c|}{0.6457}                                  & 32.25           & 29.65           \\
DCRCN~\cite{aghabiglou2021mr}                                                  & \multirow{-4}{*}{$\mathcal{C}$}                         & 0.0119          & \multicolumn{1}{c|}{0.0291}                                  & 0.9100          & \multicolumn{1}{c|}{0.8649}                                  & 32.01            & 29.49           & 0.0351          & \multicolumn{1}{c|}{0.0443}                                  & 0.7332          & \multicolumn{1}{c|}{0.6635}                                  & 32.18           & 30.76           \\ \midrule
VIT\_Base~\cite{lin2022vision}                                              &                                               & 0.0207          & \multicolumn{1}{c|}{0.0446}                                  & 0.8903          & \multicolumn{1}{c|}{0.8254}                                  & 31.33            & 28.03           & 0.0342          & \multicolumn{1}{c|}{0.0460}                                  & 0.7206          & \multicolumn{1}{c|}{0.6578}                                  & 32.10           & 30.28           \\
SwinMR~\cite{huang2022swin}                                                 &                                               & 0.0109          & \multicolumn{1}{c|}{0.0260}                                  & 0.9298          & \multicolumn{1}{c|}{0.8695}                                  & 34.14            & 30.36           & 0.0342          & \multicolumn{1}{c|}{0.0476}                                  & 0.7213          & \multicolumn{1}{c|}{0.6537}                                  & 32.14           & 30.21           \\
ReconFormer~\cite{10251064}                                            &                                               & 0.0108          & \multicolumn{1}{c|}{0.0276}                                  & 0.9297          & \multicolumn{1}{c|}{0.8650}                                  & 34.16            & 30.11           & 0.0320          & \multicolumn{1}{c|}{0.0431}                                         & 0.7327          & \multicolumn{1}{c|}{0.6672}                                        & \textbf{32.53}  & 30.76                \\
Restormer~\cite{zamir2022restormer}                                              &                                               & 0.0164          & \multicolumn{1}{c|}{0.0367}                                  & 0.9093          & \multicolumn{1}{c|}{0.8445}                                  & 32.36            & 28.86           &0.0339                 & \multicolumn{1}{c|}{0.0450}                                        & 0.7223               & \multicolumn{1}{c|}{0.6597}                                        & 32.20               &  30.46               \\
AST~\cite{zhou2024adapt}                                                    & \multirow{-5}{*}{$\mathcal{T}$}                         & 0.0149          & \multicolumn{1}{c|}{0.0322}                                  & 0.9115          & \multicolumn{1}{c|}{0.8544}                                  & 32.78            & 29.45           & 0.0335          & \multicolumn{1}{c|}{0.0445}                                  & 0.7234          & \multicolumn{1}{c|}{0.6620}                                  & 32.26           & 30.52           \\
\rowcolor[HTML]{EFEFEF} 
\textbf{Ours}                                          & \multicolumn{1}{l|}{\cellcolor[HTML]{EFEFEF}} & \textbf{0.0103} & \multicolumn{1}{c|}{\cellcolor[HTML]{EFEFEF}\textbf{0.0217}} & \textbf{0.9321} & \multicolumn{1}{c|}{\cellcolor[HTML]{EFEFEF}\textbf{0.8828}} & \textbf{34.38}   & \textbf{31.15}  & \textbf{0.0316} & \multicolumn{1}{c|}{\cellcolor[HTML]{EFEFEF}\textbf{0.0408}} & \textbf{0.7337} & \multicolumn{1}{c|}{\cellcolor[HTML]{EFEFEF}\textbf{0.6692}} & 32.51           & \textbf{31.03}  \\ \bottomrule
\end{tabular}
\vspace{-0.6 em}
    \caption{Performance comparison of MRI reconstruction under $4 \times$ and $8 \times$ Acceleration Factor (AF) on the single-coil datasets, including CC359 and fastMRI. $\mathcal{C}$: CNN-based methods. $\mathcal{T}$: transformer-based methods.}
    \label{table1}
\vspace{-1.5 em}
\end{table*}
\begin{table}[]
\renewcommand{\arraystretch}{0.8}
    \centering
    \small
    \setlength{\tabcolsep}{1.2mm}
\begin{tabular}{l|llllll}
\toprule
\multicolumn{1}{c|}{}                          & \multicolumn{6}{c}{SKM-TEA}                                                                                                                                                                               \\ \cmidrule{2-7} 
\multicolumn{1}{c|}{}                          & \multicolumn{2}{c|}{NMSE $\downarrow$}                                                & \multicolumn{2}{c|}{SSIM $\uparrow$}                                                & \multicolumn{2}{c}{PSNR $\uparrow$}                            \\ \cmidrule{2-7} 
\multicolumn{1}{l|}{\multirow{-4}{*}{Method}} & \multicolumn{1}{c}{AF=4} & \multicolumn{1}{c|}{AF=8}                     & \multicolumn{1}{c}{AF=4} & \multicolumn{1}{c|}{AF=8}                     & \multicolumn{1}{c}{AF=4} & \multicolumn{1}{c}{AF=8} \\ \midrule
KIKI-Net                                       & 0.0196                   & \multicolumn{1}{l|}{0.0271}                   & 0.8577                   & \multicolumn{1}{l|}{0.7941}                   & 34.26                    & 31.42                    \\
UNet-32                                        & 0.0204                   & \multicolumn{1}{l|}{0.0270}                   & 0.8469                   & \multicolumn{1}{l|}{0.7904}                   & 33.91                    & 31.44                    \\
D5C5                                           & 0.0188                   & \multicolumn{1}{l|}{0.0257}                   & 0.8648                   & \multicolumn{1}{l|}{0.8030}                   & 34.63                    & 31.89                    \\
SwinMR                                         & 0.0192                   & \multicolumn{1}{l|}{0.0256}                   & 0.8597                   & \multicolumn{1}{l|}{0.8022}                   & 34.45                    & 31.94                    \\
ReconFormer                                    & 0.0179                   & \multicolumn{1}{l|}{0.0239}                   & 0.8730                   & \multicolumn{1}{l|}{0.8158}                   & 35.06                    & 32.51                    \\
AST                                            & 0.0172                         & \multicolumn{1}{l|}{0.0297}                         &0.8914                          & \multicolumn{1}{l|}{0.8407}                         & 35.17                          & 32.61                         \\
\rowcolor[HTML]{EFEFEF} 
\textbf{Ours}                                           & \textbf{0.0158}                          & \multicolumn{1}{l|}{\cellcolor[HTML]{EFEFEF}\textbf{0.0200}} &\textbf{0.8975}                          & \multicolumn{1}{l|}{\cellcolor[HTML]{EFEFEF}\textbf{0.8527}} & \textbf{35.64}                         & \textbf{32.85}                         \\ \bottomrule
\end{tabular}
\vspace{-0.6 em}
    \caption{Performance comparison under $4 \times$ and $8 \times$ Acceleration Factor (AF) on the multi-coil SKM-TEA dataset.}
    \label{table2}
\vspace{-1.5 em}
\end{table}

\subsection{Hybrid Experts Feature Refinement}
Inspired by \cite{chen2023hybrid}, we introduce Hybrid Experts Feature Refinement (HEFR) to provide fine-grained information, as shown in Figure~\ref{fig:Figure2} (e). Specifically, we extract fine-grained knowledge by carefully selecting multiple CNN operations, referred to as experts. These include average pooling, separable convolution layers, and dilated convolution layers with different kernel sizes. Unlike the traditional approach of combining experts with an external gating network, we employ a self-attention mechanism as a switcher among different experts, adaptively emphasizing the importance of various feature representations based on the input. Specifically, given the input feature $F_{h} \in \mathbb{R}^{H \times W \times C}$, we first apply the channel-wise average to generate a $C$-dimensional channel descriptor $\mathcal{K} \in \mathbb{R}^{C}$:
\begin{equation}
\mathcal{K}= \textstyle{\frac{1}{H\times W}}  {\textstyle \sum_{i=1}^{H}} {\textstyle \sum_{j=1}^{W}}F_{h}(i,j)
\end{equation}
where $F_{h}(i,j)$ is the value of feature $F_{h}$ at spatial location $(i,j)$.
Then, the coefficient vector $\mathcal{V}$ of each expert is allocated corresponding to the learnable weight matrices $W_{1}\in \mathbb{R}^{D \times C}$ and $W_{2}\in \mathbb{R}^{E \times D}$, i.e., $\mathcal{V}=W_{2}\sigma(W_{1}\mathcal{K})$. Here, $D$ is the dimension of the weight matrix, $E$ is the number of experts, and $\sigma(\cdot)$ is a ReLU function. Finally, denoting the expert operations as $f_{exp}(\cdot )$, the output $F_{h}^{'}$ is obtained as:
\begin{equation}
\begin{aligned}
F_{h}^{'}=f_{1\times 1} (\textstyle \sum_{e=1}^{E} f_{exp}(F_{h},\mathcal{V})  ) +F_{h}
\end{aligned}
\end{equation}

\section{Experiments}
\subsection{Experimental Settings}
\subsubsection{Datasets and Evaluation Metrics}
The proposed FPS-Former is evaluated on three datasets: CC359 \cite{warfield2004simultaneous}, fastMRI \cite{zbontar2018fastmri}, and SKM-TEA \cite{desai2022skm}. The CC359 dataset is a publicly available raw brain MRI dataset acquired from clinical MR scanners (Discovery MR750; GE Healthcare, Waukesha, WI, USA). Following the official dataset split, we randomly selected a training set comprising $4,524$ slices from $25$ subjects, and a test set consisting of $1,700$ slices from an additional $10$ subjects. The acquisition matrix size is $256\times256$; The fastMRI dataset contains $1,172$ complex-valued single-coil coronal proton density (PD)-weighted knee MRI scans. Each scan provides approximately $35$ coronal cross-sectional knee images with the matrix of size $320\times320$. We partition this dataset into $973$ scans for training and $199$ scans (fastMRI validation dataset) for testing; The SKM-TEA raw data track provides $155$ complex-valued multi-coil T2-weighted knee MRI scans. $124, 10$, and $21$ coil-combined volumes are used for training, validation, and testing. Each subject provides approximately $160$ cross-sectional knee images with the matrix of size $512\times512$. In comparison experiments, the input under-sampled image sequences are generated by randomly under-sampling the k-space data using the 1D cartesian under-sampling function similar to the fastMRI challenge \cite{zbontar2018fastmri}. Normalized mean square error (NMSE), structural index similarity (SSIM), and peak signal-to-noise ratio (PSNR) are used as evaluation metrics for comparison.

\subsubsection{Training Details}
In our model, $(N_{0}, N_{1}, N_{2}, N_{3}, N_{4})$ are set to $(4, 1, 2, 2, 1)$, and the number of attention heads for $(N_{1}, N_{2}, N_{3}, N_{4})$ FPS blocks are set to $(1, 2, 4, 8)$. For each FPS block, the number of frequency pyramid levels $M$ in FMAM, the number of groups $N$ in SHAM, and the channel expansion factor $r$ in SDFN are set to $(3,4,2)$, respectively. For the HEFR module, we set the number of experts $E$ to $8$ and the dimension of the weight matrix $D$ to $32$. During training, we used the AdamW optimizer with a batch size of $4$ and patch size of $8$, for a total of $300K$ iterations. The initial learning rate is fixed at $1\times 10^{-4}$ for the first $92K$ iterations, then reduced to $1\times 10^{-6} $ using a cosine annealing schedule over the remaining $208K$ iterations. The entire framework is implemented on PyTorch using RTX $3090$.
\begin{figure*}[t]
    \centering
    \includegraphics[width=1.0\linewidth]{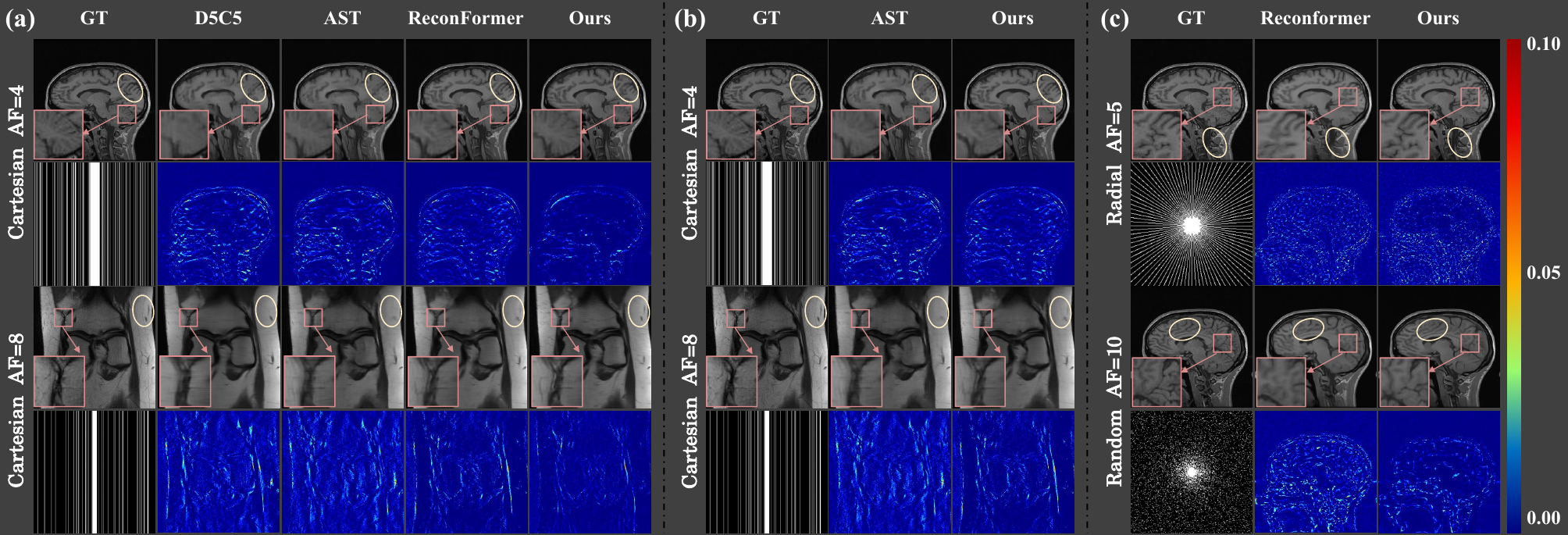}
    \caption{Qualitative comparison of different methods on (a) the single-coil dataset including CC359 and fastMRI, (b) the multi-coil dataset SKM-TEA, and (c) the CC359 dataset using different undersampling masks. The second row of each subplot shows the corresponding error maps. The red boxes and yellow ellipses highlight the details in the reconstruction results.}
    \label{fig:Figure3}
    \vspace{-1.0 em}
\end{figure*}

\subsection{Comparison with State-of-the-arts}
\paragraph{Single-coil datasets}
We compared the proposed FPS-Former with recent MRI reconstruction approaches, including CNN-based and Transformer-based methods. Additionally, we evaluated it against two state-of-the-art natural image restoration methods, Restormer \cite{zamir2022restormer} and AST \cite{zhou2024adapt}, which were equipped with a DC layer with the same settings of MRI reconstruction for a fair comparison. Table~\ref{table1} shows the comparison results of our FPS-Former with other methods under different acceleration factors (AF) on single-coil datasets, including CC359 and fastMRI. As shown in this table, FPS-Former demonstrates significant improvements over CNN-based methods and consistently surpasses other Transformer-based approaches across different acceleration rates on both datasets. For example, our method shows the superiority of $2.37$ dB over the CNN-based SoTA DCRCN and $0.22$ dB over Transformer-based counterpart ReconFormer under $4 \times$ AF on CC359. Notably, our approach shows greater performance improvement as the acceleration factor increases, particularly in more challenging scenarios. Specifically, for the CC359 and fastMRI datasets, our model outperforms the leading method AST, by $1.70$ dB and $0.51$ dB at $8 \times$ AF, and $1.60$ dB and $0.25$ dB at $4 \times$ AF, respectively.

\paragraph{Multi-coil datasets}
Table~\ref{table2} gives comparison results of MRI reconstruction on the multi-coil SKM-TEA dataset. We achieved $35.64$ and $32.85$ PSNR under $4 \times$ and $8 \times$ AF respectively. Our FPS-Former significantly outperforms previous CNN-based solutions and shows the superiority of $0.47$ dB and $0.24$ dB over AST at $4 \times$ AF and $8 \times$ AF, respectively. This further demonstrates the superiority of our method.

\begin{table}[t]
\renewcommand{\arraystretch}{0.9}
    \centering
    \small
    \setlength{\tabcolsep}{0.4 mm}
\begin{tabular}{l|c|ll|ll|ll}
\toprule
                         &                        & \multicolumn{2}{c|}{NMSE $\downarrow$}                                & \multicolumn{2}{c|}{SSIM $\uparrow$}                                & \multicolumn{2}{c}{PSNR $\uparrow$}                              \\
\multirow{-2}{*}{Method} & \multirow{-2}{*}{Mask} & \multicolumn{1}{c}{AF=5}   & \multicolumn{1}{c|}{AF=10}  & \multicolumn{1}{c}{AF=5}   & \multicolumn{1}{c|}{AF=10}  & \multicolumn{1}{c}{AF=5}  & \multicolumn{1}{c}{AF=10} \\ \midrule
ReconFormer              &                        & \multicolumn{1}{c}{0.0070} & \multicolumn{1}{c|}{0.0170} & \multicolumn{1}{c}{0.9450} & \multicolumn{1}{c|}{0.8971} & \multicolumn{1}{c}{36.18} & \multicolumn{1}{c}{32.17} \\
AST                      & \multirow{-2}{*}{$\mathcal{I}$}   &\multicolumn{1}{c}{0.0068}                 & \multicolumn{1}{c|}{0.0173}  &  \multicolumn{1}{c}{0.9448}                 &\multicolumn{1}{c|}{0.8988}            & \multicolumn{1}{c}{36.12}                          &\multicolumn{1}{c}{32.10}                           \\
\rowcolor[HTML]{EFEFEF} 
\textbf{Ours}                     &                        & \multicolumn{1}{c}{\textbf{0.0060}}  & \multicolumn{1}{c|}{\textbf{0.0163}}          & \multicolumn{1}{c}{\textbf{0.9467}}  & \multicolumn{1}{c|}{\textbf{0.9017}}    &\multicolumn{1}{c}{\textbf{36.40}}  & \multicolumn{1}{c}{\textbf{32.36}}                           \\ \midrule
ReconFormer              &                        & \multicolumn{1}{c}{0.0125} & \multicolumn{1}{c|}{0.0176} & \multicolumn{1}{c}{0.9164} & \multicolumn{1}{c|}{0.8918} & \multicolumn{1}{c}{33.52} & \multicolumn{1}{c}{32.02} \\
AST                      & \multirow{-2}{*}{$\mathcal{R}$}   & 0.0127                        & \multicolumn{1}{c|}{0.0178}                             & 0.9142                     &\multicolumn{1}{c|}{0.8924}                             & 33.48                      & \multicolumn{1}{c}{31.98}                           \\
\rowcolor[HTML]{EFEFEF} 
\textbf{Ours}                     &                        & \multicolumn{1}{c}{\textbf{0.0119}}           & \multicolumn{1}{c|}{\textbf{0.0173}}                             & \textbf{0.9170}            & \textbf{0.8943}                             & \multicolumn{1}{c}{\textbf{33.74}}             & \multicolumn{1}{c}{\textbf{32.11}}                           \\ \bottomrule
\end{tabular}
\vspace{-0.6 em}
    \caption{Performance comparison of MRI reconstruction under $5 \times$ and $10 \times$ Acceleration Factors (AF) on the CC359 using more masks. $\mathcal{I}$: Radial mask. $\mathcal{R}$: Random mask.}
    \label{table3}
\vspace{-1.0 em}
\end{table}
\begin{table}[t]
\renewcommand{\arraystretch}{0.9}
    \centering
    \small
    \setlength{\tabcolsep}{1.0mm}
\begin{tabular}{l|ccccccc}
\toprule
Model         & FMAM & SPAM & SDFN & HEFR & NMSE            & SSIM             & PSNR           \\ \midrule
(a)           &      & \pmb{$\checkmark$}    & \pmb{$\checkmark$}     &\pmb{$\checkmark$}    & 0.0109          & 0.9294          & 34.14          \\
(b)           & \pmb{$\checkmark$}    &      & \pmb{$\checkmark$}     & \pmb{$\checkmark$}    & 0.0117          & 0.9246          & 33.83          \\
(c)           & \pmb{$\checkmark$}    & \pmb{$\checkmark$}    &       & \pmb{$\checkmark$}    & 0.0114          & 0.9260          & 33.95          \\
(d)           & \pmb{$\checkmark$}    & \pmb{$\checkmark$}    & \pmb{$\checkmark$}     &      & 0.0113          & 0.9274          & 33.98          \\
\rowcolor[HTML]{EFEFEF} 
\textbf{Ours} & \pmb{$\checkmark$}    & \pmb{$\checkmark$}    & \pmb{$\checkmark$}     & \pmb{$\checkmark$}    & \textbf{0.0103} & \textbf{0.9321} & \textbf{34.38} \\ \bottomrule
\end{tabular}
\vspace{-0.6 em}
    \caption{Ablation results on FPS-Former on CC359 (AF=$4$).}
    \label{table4}
\vspace{-2.0 em}
\end{table}
\paragraph{Experiments on different masks}
To further demonstrate the robustness of our FPS-Former, we conducted experiments using radial and random undersampling patterns under $5 \times$ and $10 \times$ acceleration factors on CC359 dataset. As shown in Table~\ref{table3}, FPS-Former consistently outperforms other methods, highlighting its ability to effectively reconstruct MRI images from various undersampling masks.

\paragraph{Visualization Results}
The qualitative results for the single-coil dataset, multi-coil dataset, and mask experiments are shown in Figure \ref{fig:Figure3} (a), (b), and (c), respectively. In (a), the CNN-based SoTA, D5C5, suffers from severe edge blurring and substantial detail loss. Although the Transformer-based AST and ReconFormer partially alleviate these issues, they still lose crucial anatomical details in challenging tasks with high acceleration factors. In contrast, our FPS-Former demonstrates robustness to various anatomical structures and acceleration factors. By addressing the issues of ViT models, our method better preserves important anatomical details, as highlighted by the zoomed boxes and ellipses. In (b), our FPS-Former can restore more abundant details than other counterparts. This further demonstrates that our method can effectively reconstruct not only single-coil but also multi-coil MRI images. In (c), our method demonstrates great robustness across various undersampling patterns and acceleration rates, further validating its effectiveness.

\subsection{Ablation Studies and Analysis}
\subsubsection{Efficacy of Key Components}
We first performed a break-down ablation to investigate the effect of each component and their interactions. As the results listed in Table~\ref{table4} show, (a) When FMAM is removed, the performance dramatically degrades by $0.24$ in PSNR and $0.0027$ in SSIM. This drop can be attributed to FMAM's ability to preserve high-frequency details, which are crucial for restoring local textures and edges. (b) Excluding SPAM leads to significant reductions in PSNR and SSIM by $0.55$ and $0.0075$, respectively. This demonstrates that SPAM effectively mitigates the noise impact from content-irrelevant tokens, thereby enhancing performance. (c) Replacing SDFN with a conventional feed-forward network in the standard ViT results in a decrease in PSNR from $34.38$ to $33.95$. This highlights SDFN's effectiveness in representing multi-scale features, which is essential for improved MRI reconstruction. (d) The absence of HEFR results in a decline of $0.40$ in PSNR and $0.0047$ in SSIM, showing its significant contribution. 

\begin{figure}[t]
    \centering
    \includegraphics[width=1.0\linewidth]{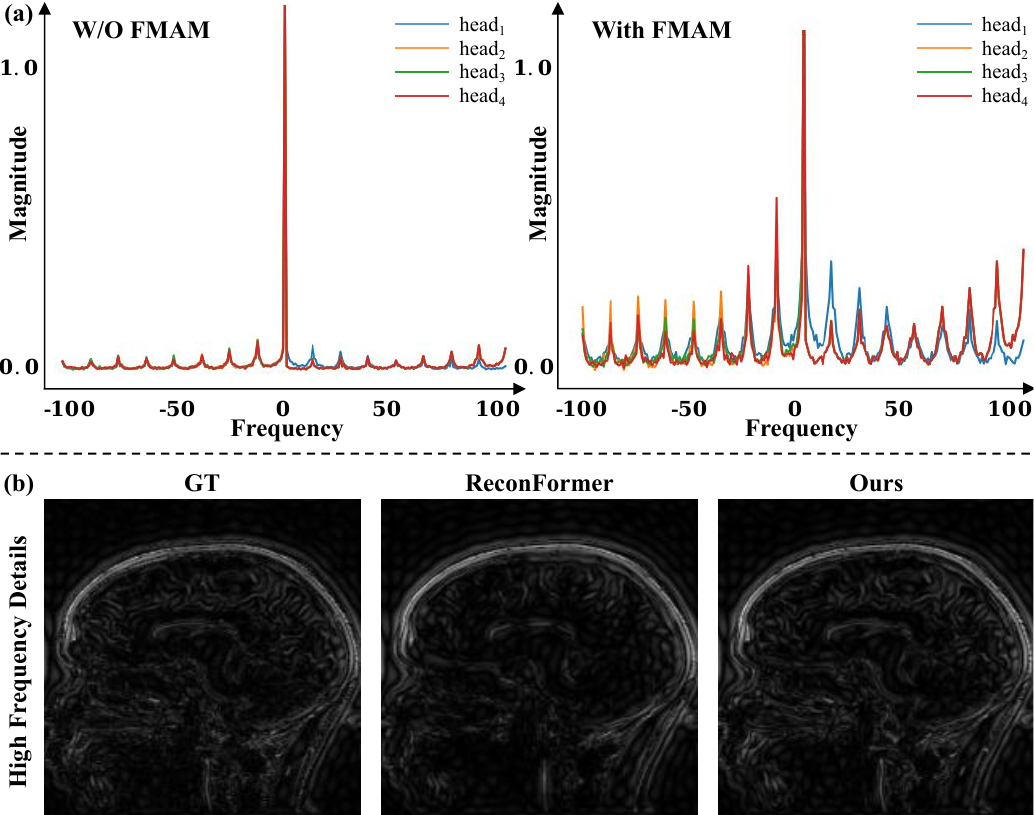}
    \caption{(a) Frequency response analysis. (b) Visualization results of high-frequency details in reconstructed images.}
    \label{fig:Figure4}
    \vspace{-1.5 em}
\end{figure}

\subsubsection{Analysis of FPS block}
As shown in Table~\ref{table5}, to further analyze the effectiveness of our FPS block, we compared various variants for FMAM, SPAM, and SDFN. Specifically, (1) we investigated the impact of the number of frequency pyramid layers $M$ in FMAM. Our results indicate that both too small and large $M$ negatively affect network performance, with the optimal reconstruction results achieved when $M=3$. Fewer layers struggle to capture diverse frequency information, while excessive layers lead to confusion in feature aggregation. (2) We compared our SPAM with several self-attention mechanisms, including global MSA and local window-based MSA. We can see that, SPAM shows the most significant improvement because it performs MSA calculations among tokens with closely related content, effectively reducing noisy interactions. (3) We evaluated SDFN against three baseline methods: the conventional feed-forward network (FN), the depth-wise convolution feed-forward network (DFN), and the gated-depth-wise convolution feed-forward network (GDFN). Although GDFN employs a gating mechanism to enhance performance, it does not leverage multi-scale feature integration, which is crucial for MRI reconstruction. In contrast, our SDFN incorporates local feature extraction and fusion across different scales, achieving a PSNR gain of $0.56$ dB over GDFN.

\subsubsection{Efficiency of Frequency Modulation Attention Module}
To further validate the effectiveness of our FMAM, we follow \cite{wang2022anti} and present a spectral response comparison between network variants with and without FMAM at the last encoder layer, as shown in Figure \ref{fig:Figure4} (a). The frequency response of the network without FMAM exhibits greater attenuation of high frequency compared to FPS-Former. Additionally, we extracted high-frequency structures from the reconstructed images of different methods using a high-pass filter, as illustrated in Figure \ref{fig:Figure4} (b). Due to FMAM's effective preservation of high-frequency details, our method reconstructs edges and textures more accurately and completely. This visual evidence underscores FMAM's superior ability to tackle ViT's low-pass filter issues.

\subsubsection{Analysis of Training Efficiency}
The training efficiency comparison is reported in Table~\ref{table6}. The recent ViT-based ReconFormer employs a recurrent structure to maintain a few trainable parameters. However, its significantly higher computational complexity (FLOPs=$342$G) substantially increases both training difficulty and inference time. On the other hand, AST addresses both spatial and channel redundancy and achieves a notable reduction in FLOPs. Nevertheless, AST has a larger parameter count of 26.10 M and a noticeable drop in performance. Compared to these methods, our FPS-Former achieves significant performance improvements while maintaining both low computational complexity and a minimal number of trainable parameters. This advantage is largely attributed to our SPAM, which minimizes interactions between irrelevant tokens, effectively reducing both noise interference and computational complexity.

\subsubsection{Analysis of Hyper-parameters}
The analysis of key hyper-parameters, such as the number of experts in HEFR, the number of frequency pyramid levels in FMAM, the expansion ratio in SDFN, and the number of FPS blocks and HEFR modules, etc., is specifically discussed in the \textit{Supplementary Materials}. FPS-Former demonstrates consistent performance across different hyper-parameter variations.

\begin{table}[t]
\renewcommand{\arraystretch}{0.8}
    \centering
    \small
    \setlength{\tabcolsep}{1.2mm}
\begin{tabular}{lc|ccc}
\toprule
\multicolumn{1}{l|}{Model}                       & Com.                             & NMSE            & SSIM            & PSNR           \\ \midrule
\multicolumn{1}{l|}{2 layered pyramid}               &                                  & 0.0104          & 0.9312          & 34.30          \\
\multicolumn{1}{l|}{4 layered pyramid}               & \multirow{-2}{*}{$\mathcal{F}$}              & 0.0103          & 0.9320          & 34.36          \\ \midrule
\multicolumn{1}{l|}{Global-MSA~\cite{alexey2020image}}                      &                                  & 0.0117          & 0.9246          & 33.83          \\
\multicolumn{1}{l|}{Window-MSA~\cite{liu2021swin}}                      & \multirow{-2}{*}{$\mathcal{P}$}              & 0.0115          & 0.9261          & 33.89          \\ \midrule
\multicolumn{1}{l|}{FN~\cite{alexey2020image}}                          &                                  & 0.0114          & 0.9260          & 33.95          \\
\multicolumn{1}{l|}{DFN~\cite{li2021localvit}}                         &                                  & 0.0111          & 0.9277          & 34.05          \\
\multicolumn{1}{l|}{GDFN~\cite{zamir2022restormer}}                        & \multirow{-3}{*}{$\mathcal{S}$}              & 0.0117          & 0.9247          & 33.82          \\ \midrule
\rowcolor[HTML]{EFEFEF} 
\multicolumn{2}{l|}{\cellcolor[HTML]{EFEFEF}\textbf{Ours} (3 layered+SPAM+SDFN)} & \textbf{0.0103} & \textbf{0.9321} & \textbf{34.38} \\ \bottomrule
\end{tabular}
\vspace{-0.6 em}
    \caption{Ablation study for variants of FMAM ($\mathcal{F}$), SPAM ($\mathcal{P}$), and SDFN ($\mathcal{S}$) on CC359 under $4 \times$ acceleration factor.}
    \label{table5}
\vspace{-0.9 em}
\end{table}
\begin{table}[t]
\renewcommand{\arraystretch}{1.07}
    \centering
    \small
    \setlength{\tabcolsep}{0.9mm}
\begin{tabular}{l|cccc}
\toprule
\multicolumn{1}{l|}{Method} & FLOPs & Param. & SSIM            & PSNR           \\ \midrule
ReconFormer~\cite{10251064}                 & 342G  & 1.14M  & 0.9297          & 34.16          \\
AST~\cite{zhou2024adapt}                         & 155G  & 26.10M & 0.9115          & 32.78          \\
\textbf{Ours}               & 152G  & 12.51M & \textbf{0.9321} & \textbf{34.38} \\ \bottomrule
\end{tabular}
\vspace{-0.6 em}
    \caption{Efficiency of FPS-Former compared to others. The experiment is conducted on CC359 (AF=$4$) with RTX $3090$.}
    \label{table6}
\vspace{-2.0 em}
\end{table}

\section{Conclusion}
In this work, we propose to boost ViT-based MRI reconstruction by tackling three issues, including loss of high-frequency information, redundancy interactions among irrelevant tokens, and challenges in multi-scale feature modeling. To achieve this, we propose the frequency modulation attention module for frequency information correction, the spatial purification attention module for grouped token interactions, and the scale diversification feed-forward network for multi-scale feature transmission, respectively. Extensive experiments and analysis are conducted on CC359, fastMRI, and SKM-TEA datasets, validating the efficiency of FPS-Former in tackling the issues of ViT-MRI and significantly improving the performance of MRI reconstruction.

\section{Acknowledgments}
This work was supported by the National Natural Science Foundation of China under Grant No.82372097.

\bibliography{aaai25}

\end{document}